\documentclass[runningheads]{llncs}

\usepackage{graphicx}
\usepackage{makecell}
\usepackage{adjustbox}
\usepackage[hyphens]{url}
\usepackage[hidelinks]{hyperref}
\hypersetup{breaklinks=true}

\begin{document}
\title{Test-driving RISC-V Vector hardware for HPC}
\author{Joseph K. L. Lee\inst{1}\orcidID{0000-0002-1648-2740} \and
Maurice Jamieson\inst{1}\orcidID{0000-0003-1626-4871} \and
Nick Brown\inst{1}\orcidID{0000-0003-2925-7275} \and Ricardo Jesus\inst{1}\orcidID{0000-0002-9651-4756}}
\authorrunning{J. K. L. Lee et al.}
\institute{EPCC, University of Edinburgh, Bayes Centre, 47 Potterrow, Edinburgh, United Kingdom\\
\email{\{j.lee,m.jamieson,n.brown\}@epcc.ed.ac.uk}, \email{rjj@ed.ac.uk}
}
\maketitle              % typeset the header of the contribution
\begin{abstract}
Whilst the RISC-V Vector extension (RVV) has been ratified, at the time of writing both hardware implementations and open source software support are still limited for vectorisation on RISC-V. This is important because vectorisation is crucial to obtaining good performance for High Performance Computing (HPC) workloads and, as of April 2023, the Allwinner D1 SoC, containing the XuanTie C906 processor, is the only mass-produced and commercially available hardware supporting RVV. This paper surveys the current state of RISC-V vectorisation as of 2023, reporting the landscape of both the hardware and software ecosystem. Driving our discussion from experiences in setting up the Allwinner D1 as part of the EPCC RISC-V testbed, we report the results of benchmarking the Allwinner D1 using the RAJA Performance Suite, which demonstrated reasonable vectorisation speedup using vendor-provided compiler, as well as favourable performance compared to the StarFive VisionFive V2 with SiFive's U74 processor. 
\end{abstract}

\section{Introduction}

Vector instructions bring many benefits to an Instruction Set Architecture (ISA), for instance they enable applications to exploit data parallelism, reduce code size, increase instruction bandwidth and improve energy efficiency. Many modern applications including machine learning, graphics, digital signal processing, and cryptography are built around algorithms that are designed to heavily take advantage of vector instructions. Indeed vectorisation was a traditional way in which HPC was undertaken on the likes of the Cray-1 and Thinking Machines' CM series before distributed memory parallelism became widespread. Modern day variants of these ideas, such as AVX-512, the NEC SX-Aurora Vector Engine and the flexibility provided by Arm SVE in the A64FX, are highly successful.

Over the past years RISC-V has become a well-established open ISA standard, where RISC-V is the fifth major RISC ISA design from the Univerity of California Berkeley, preceded by RISC-I, RISC-II, SOAR, and SPUR. The most powerful feature of RISC-V in comparison to other RISC designs, such as the SPARC, PowerPC, MIPS and Arm, is its modular design. In practice this means that a small base integer ISA is specified and then ISA extensions, such as floating-point and vector support, can be chosen and added to the CPU implementation. Vector support has been a key extension for RISC-V since its inception, \emph{We also pun on the use of the Roman numeral ``V'' to signify ``variations'' and ``vectors'', as support for a range of architecture research, including various data-parallel accelerators, is an explicit goal of the ISA design.} \cite{riscv-instruction-set-manual-2019}
%% MJ: Not sure we need this statement
%% The RISC-V ISA was born to investigate vector engines, so there should be no question that a robust vector extension should exist on top of the base ISA.

Version 1.0 \cite{risc-v-v-extension-v1p0} of the RISC-V vector extension (RVV) was ratified in late 2021. Similarly to Arm SVE, it is inherently vector length agnostic (VLA) and the same code can be executed on implementations with different vector lengths, and the element size and vector length can also be reconfigured at run time. Whereas the x86 AVX and Arm NEON use the vector length specific (VLS) approach of packed SIMD and the code will need to be re-optimised and re-compiled for each vector processor, VLA code remains portable across different vector processor design and generations. 
%% MJ: I'm not sure what we mean here - shouldn't we mention the D1 boards, then call out this benchmark? 

RVV has already been used in production for physical RISC-V hardware, for example T-Head's XuanTie C906 core provides RVV v0.7.1 and has made a submission for MLPerf Tiny Inference\cite{noauthor_v07_2022}, a benchmark designed to measure trained neural network performance for low power devices. However, as an emerging standard it is not entirely straightforward to utilise and test the RISC-V vector extension. This paper aims to evaluate the current landscape when it comes to RISC-V vectorisation and assess the potential gain from utilising RISC-V vectors for HPC applications. Ultimately our objective is to provide guidance for users interested in testing or adopting available vector hardware using experiences we have gained from setting up the EPCC RISC-V testbed\cite{team_excalibur_nodate}. The key contributions of this paper are:

\begin{enumerate}
    \item  We review the state of play of the RISC-V vector extension and available processor implementations
    \item We evaluate the availability of open source software such as compiler toolchains and Linux kernels to support running vectorised code on available hardware
    \item We perform benchmarks and evaluate vectorisation efficiency using a currently available compiler and commercially available RISC-V vector processor
\end{enumerate}

\section{Background and related work}
\subsection{V Extension}
The RISC-V `V' standard extension introduces 32 new vector registers, and requires a minimum vector register length (VLEN) of 128 bits up to a maximum 65,536 bits.\footnote{The Zvl32b and Zvl64b extensions allow for a smaller minimum VLEN of 32 and 64 bits respectively} This can be compared to SVE, which also has a minimum vector length of 128 bits, but only a maximum of 2048 bits. Another feature of the vector instruction set is that multiple vector registers can be grouped together as a single combined vector and this is known as \emph{LMUL}. Whilst previously one could only group 2, 4 or 8 registers, in RVV v1.0 fractional groupings of $\frac{1}{2}$, $\frac{1}{4}$ and $\frac{1}{8}$ are also allowed where part of a single vector register will be used. These features of the instruction set provide great flexibility because, within a single code, the vector length can be varied by different groupings of vector registers dynamically, and this is therefore particularly useful when operating on mixed-width values. Combined with the fact that the same compiled code can run on hardware implementations with significantly different vector width and automatically exploit the widest vector lengths, RVV encourages portable code with greater utilisation of vector register resources without the need for platform-specific optimisation.

Prior to the ratification of v1.0 of the V extension, the beta version of RVV, v0.7.1, was adopted in production for example by the XuanTie C906 processor and BSC's Vitruvius+\cite{minervini_vitruvius_2022} which is part of the European Processor Initiative (EPI) project. Even though the difference between the v1.0 and v0.7.1 is fairly minimal, the two versions are incompatible in terms of source code or binary. One major difference is the lack of support for fractional LMUL in version 0.7.1. %We have developed a tool to backport version 1.0 assembly to version 0.7.1, which is introduced in a separate paper for this workshop. The two versions are compared in more detail, and the auto-vectorization capabilities and performance of available compilers targeting v0.7.1 and v1.0 are evaluated.

\subsection{Intrinsics}
At the time of writing, the official RISC-V task group is converging towards v1.0 of the C intrinsics API~\cite{noauthor_risc-v_2023}, which is expected to be released later in 2023. Currently, LLVM supports v0.10 of the intrinsics specification and mainline GCC provides no support at all. It is in the roadmap of both compilers to support v1.0 in the future once it is ratified. However the XuanTie 900 series toolchain, which is a modified version of the GCC 8.4 compiler targeting the C906 and C910 supports a custom set of intrinsics for v0.7.1 and v1.0. As does the LLVM compiler from BSC for the EPI project's RISC-V Toolchain~\cite{noauthor_bsc_2022} providing their set of v0.7.1 and v1.0 intrinsics. These bespoke compiler versions can be useful when developing for vectorisation due to limitations in the mainline compilers. 

\subsection{P Extension}
It should also be noted that there is packed a SIMD `P' extension to the base ISA which uses the floating point registers and is aimed at embedded cores and low-power digital signal processing (DSP) applications, such as audio and video encoding/decoding, image interpretation and computer vision. The extension has not yet been ratified, the latest version is v0.9.11 \cite{noauthor_riscv-p-specp-ext-proposalpdf_nodate}, and provides a large number of SIMD and partial-SIMD instructions, such as 8/16-bit minimum and maximum instructions (including \texttt{SMIN8}, \texttt{UMIN8}, \texttt{SMAX16} and \texttt{UMAX16}), and 16/32-bit multiply with 64-bit add/subtract instructions (including \texttt{SMAL}, \texttt{SMALBB} and \texttt{SMAR64}).

\subsection{Related work}
Even though RVV has been ratified relatively recently, studies focusing on other (scalable) vector ISAs can be applicable when wishing to improve vector performance for RISC-V. For example, there has been studies comparing the performance of Arm SVE against NEON \cite{pohl_performance_2019} and AVX \cite{soria-pardos_use_2021}, and evaluating the vectorisation efficiency and usage on mini-apps for available SVE compilers \cite{poenaru_evaluating_2020}. Another parameter which has significant impact on performance with the VLA programming model is the implementation vector length, where \cite{odajima_performance_2021} and \cite{ramirez_risc-v_2020} study the performance of a variety of vectorised applications with different vector lengths using the gem5 simulator for Arm SVE and RVV respectively.

There is currently a rapid development of research-based RVV enabled hardware underway, for example ETH Zurich have introduced \textit{Ara} \cite{Ara2020} and its upgrade \cite{perotti_new_2022}, and BSC introduced \textit{Vitruvius+} \cite{minervini_vitruvius_2022}. Whilst none are yet mass-produced or widely available, these RISC-V vector accelerator designs have been taped-out and their performance compared in \cite{minervini_vitruvius_2022}.

\section{RVV CPU Implementations}

There is a broad selection of IP cores which have implemented RVV and this is summarised in Table~\ref{tab:IP-cores}. RISC-V cores on this list target a wide range of applications, including edge artificial intelligence/machine learning (SiFive X280), general high-performance application (SiFive P series), and decoupled vector accelerator (Ara/Vitruvius+). The decoupled accelerator approach is especially interesting because this allows vector instructions to be offloaded from the scalar pipeline, and paired with support for long vectors, for instance 256 double precision elements per vector register are supported by the Vitruvius+, these present high performance RISC-V vector accelerators for HPC workloads. In taped-out implementations the New Ara core reports achieving 37.1 GFLOPS per Watt ~\cite{perotti_new_2022} and Vitruvius+ reports 47.3 GFLOPS per Watt~\cite{minervini_vitruvius_2022} on matrix multiplication benchmarks.

\begin{table}
\caption{List of available RVV processors. The last three entries are open source.}\label{tab:IP-cores}
\begin{adjustbox}{width={\textwidth},totalheight={\textheight},keepaspectratio}
\begin{tabular}{|c|c|c|}
\hline
\textbf{Processor} &  \textbf{Vector Length} & \textbf{RVV version}\\
\hline \hline
SiFive P270/P470/P670~\cite{noauthor_sifive_nodate}& 256-bit/128-bit/dual 128-bit & 1.0\\
SiFive X280~\cite{noauthor_sifive_nodate-1}& 512-bit & 1.0\\
Andes NX27V~\cite{noauthor_risc-vnx27v_nodate} & Configurable from 128 to 512-bit & 1.0\\
Andes AX45MPV~\cite{noauthor_risc-v_nodate} & Configurable from 128 to 1024-bit & 1.0\\
Vitruvius+~\cite{minervini_vitruvius_2022} & 16384-bit & 0.7.1 (update to 1.0 in future)\\
Hwacha~\cite{schmidt_eight-core_2022} (V4~\cite{schmidt_hwacha_nodate}) & 512-bit & custom\\
New Ara~\cite{perotti_new_2022} & Configurable e.g. 4096-bit& 1.0\\
Tenstorrent BOOM-ocelot~\cite{noauthor_ocelot_2023}& Configurable from 128-bit & 1.0\\
T-Head XuanTie C906~\cite{noauthor_ip_2023} & 128-bit & 0.7.1\\
\hline
\end{tabular}
\end{adjustbox}
\end{table}

These energy efficiency numbers delivered by the New Ara and Vitruvius+ cores are impressive, especially considering that they are still research prototypes rather than production parts. For comparison, whilst the Green 500 reports whole systems rather than the individual machine components, based on the November 2022 list those HPC machines that are able to achieve greater than 50 GFLOPS per Watt are based around either the AMD Instinct or Nvidia Grace Hopper GPUs. These represent mature technologies with a rich lineage, whereas by comparison the New Ara and Vitruvius+ are the first generation of RISC-V vector accelerators and therefore as time progresses are likely to significantly increase their performance and energy efficiency.

\subsubsection{Physical cores}
At the time of writing, the only mass-produced and commercially available physical RISC-V vector core is the XuanTie C906 from T-Head, which is the chip division of Alibaba. This contains 128-bit wide vector registers, and supports vector element sizes of 8, 16, and 32 bits. Noticeable by its absence however is support for elements of size 64 bits, meaning that the XuanTie C906 does not support 64 bit double precision floating point. This is a major disadvantage for HPC, where the vast majority of our workloads are in double precision. Nevertheless it is still interesting to benchmark with single precision workloads as understanding the performance and software ecosystem can provide insights around RVV, albeit at single precision. The XuanTie C906 core is available as part of the Allwinner D1 SoC, part of the EPCC RISC-V testbed and the main system on which we perform our vector benchmarks in Section \ref{sec:benchmarks}. 

\section{Toolchain and software support}
In this section we review the current status of the RISC-V open source software ecosystem which supports compiling and running vectorised code on RVV processors.

\subsection{Compiler toolchain}
\subsubsection{GNU}
At the time of writing, the upstream GNU compiler toolchain does not support the vector extension. There is a branch, \emph{rvv-next} \cite{rvv-next}, which provides limited support for RVV v1.0 and an older deleted branch \emph{rvv-0.7.1} which targeted RVV v0.7.1. T-head provides a modified GNU toolchain which targets their C906 CPU~\cite{noauthor_thead_download_nodate}, and contains optimised vectorisation for v0.7.1. This is the compiler used in this paper to benchmark the C906 CPU. Since the compiler is optimised for the C906, it generates code specifically for 128-bit vector width.

However, it should be noted that in recent weeks the T-Head GNU compiler has been removed from their download page and-so is no longer available. Because the compiler is under the GNU licence, it has been mirrored at~\cite{team_excalibur_nodate}.

\subsubsection{LLVM}
LLVM 15 and 16 support RVV v1.0, and several of the auto-vectorisation characteristics have been studied in \cite{adit_performance_2022}. LLVM supports compiling vector length agnostic RVV code via the \emph{scalable-vectorization=on} flag, as well as vector length specific via the \emph{riscv-v-vector-bits-min=N} flag (where \emph{N} is the fixed vector width in bits). LLVM also supports standard extensions with minimum vector length \emph{Zvl*} and its counterpart for embedded processors \emph{Zve*}. Since LLVM only targets RVV v1.0 and cannot run natively on the physical hardware available, it is not tested in this paper. A rollback tool that translates generated RVV v1.0 to v0.7.1 has been developed and is reported, along with a performance comparison against GCC, in~\cite{jkll_riscv_vector_tool_2023} for both VLS and VLA modes.

\subsection{Linux kernel}

Whilst there is now general availability of common Linux distributions for RISC-V boards, including Debian, Ubuntu and Fedora \cite{noauthor_architecturesrisc-vinstalling_nodate}, many are early developer variants \cite{ubuntu_download_nodate} or unsupported releases \cite{noauthor_architecturesrisc-vallwinner_nodate}. The Sipeed Linux image for the Allwinner D1, is easy to deploy using the proprietary tools and supports vectorisation out of the box. However, due to the proprietary, protected format of the bootloader, Linux images must be built using cross-compilation tools on another host and vendor-specific patches must be applied to \emph{buildroot}. Furthermore, the T-Head specific GCC compiler version must also be used for this to ensure that the resulting image is RVV compatible. 

This requirement to rebuild the bootloader and apply vendor patches is not only time consuming but also requires considerable knowledge and expertise to achieve. This is definitely an area in which the vendors of these boards could improve upon to open up their systems further and lower the barrier to entry.

%\ifx\mojocomments
%\fbox{\parbox{0.9\textwidth}{\textbf{MJ:} Talk generally about RISC-V distribution %availability, then the specifics regarding \texttt{rvv} kernel support for D1 / C906. 
%It'll be good to add in link to patched \texttt{buildroot} source to generate \texttt{rvv} kernels. }}
%\fi 

\subsection{Performance analysis tooling and instrumentation}
The RISC-V hardware ecosystem is moving very quickly and the HiFive Unmatched, released in late May 2021, and Allwinner Nezha D1, released in April 2021, are an example of where the software support sometimes struggles to keep up, especially when board and/or CPU specific support is required by tooling. Profiling tools are an example of this problem, where support for tools such as \emph{perf} has lagged the hardware. 

For instance, with the HiFive Unmatched, the Linux kernel version 5.18 only supports instruction and cycle count hardware events for \emph{perf}, and in order to obtain further events then one must patch the kernel and OpenSBI \cite{noauthor_how_perf_2022}. With the Allwinner D1, containing the XuanTie C906 core, official support for \emph{perf} was only released in the Linux kernel version 6.2 on February 19th, 2023, almost two years after the hardware was made available.

This lack of performance analysis tooling is a major drawback for HPC workloads, where it is imperative that programmers can gain insights around performance bottlenecks in codes and use this feedback to then optimise their applications.

\subsection{Emulation}
Given the limited physical hardware currently available that supports RVV, and none that supports v1.0, an obvious alternative is to run RVV-based codes under emulation. There are two main emulators for RISC-V, QEMU and Spike. Current upstream QEMU supports RVV v1.0 along with the \emph{zve32f} and \emph{zve64f} standards which provide 32-bit and 64-bit vectorisation floating point support for embedded RISC-V CPUs respectively. Versions of QEMU prior to December 20th, 2021 supported RVV v0.7.1 only. 

Likewise, Spike also supports RVV v1.0 and releases prior to November 12th, 2019 support v0.7.1. However, whilst emulation might appear to be a good choice for those wishing to experiment with RISC-V vectorisation in their applications, in absolute terms the application will run far slower than on physical hardware. Even for exploratory purposes this could be an issue as it will potentially limit the scale of testcases that can be executed.

\subsubsection{Vehave}
Developed by Barcelona Supercomputing Center (BSC), Vehave~\cite{noauthor_vehave_2021} is a functional emulator based on QEMU which is able to dynamically handle and emulate vector instructions when running vectorised binaries on hardware that does not support the vector extension. There are separate versions supporting RVV v1.0 and v0.7. Whilst this provides a convenient way of supporting RVV on hardware that is not equipped with this RISC-V extension, it is far slower than the performance that would be provided by a physical CPU.

\subsection{Softcores}
Whilst the C906 is the only RVV hard CPU core readily available, there are a number of RVV softcores, such as the Andes NX27V~\cite{noauthor_risc-vnx27v_nodate}, Andes AX45MPV~\cite{noauthor_risc-v_nodate} and Tenstorrent BOOM-ocelot~\cite{noauthor_ocelot_2023} that can be included in field-programmable gate array (FPGA) designs to test RISC-V vectorisation codes. However, creating soft-core FPGA designs requires comprehensive knowledge of the FPGA tooling and logic circuit design, such as \emph{negative slack} \cite{noauthor_clock_timing_nodate}.

\subsection{Libraries}
Most HPC libraries can be cross-compiled for RISC-V, but there tend to be limited vectorisation optimisation applied within these. One library which already includes vector optmisation is OpenBLAS, which has been optimised for RVV v0.7.1 (specifically for XuanTie C906/C910)\cite{xianyi_openblas_2023}. At the time of writing there are numerous efforts on-going across the community to optimise HPC libraries for RVV, and within the next year we will likely see significantly increased support in this regard.

\section{Benchmarks}
\label{sec:benchmarks}
\subsection{System}
The main RISC-V system that we benchmark in this paper is the Allwinner D1, which contains a C906 processor and supports RVV v0.7.1 with 128-bit vector registers. For comparison against a scalar-only RISC-V CPU, we use the StarFive VisionFive V2 board (VF2), which contains a StarFive JH7110 processor (quad core SiFive U74). In order to provide some context with similar vector designs already in use for HPC, we also performed runs on a Fujitsu A64FX system (Armv8), which supports fixed length SIMD (NEON), as well as vector length agnostic (SVE), instruction sets. 

Because the C906 only contains a single core, all benchmarks are run on a single core to enable direct comparison across CPUs, and only NEON with 128-bit vector width is used on A64FX for an objective evaluation (the XuanTie GCC compiler only generates fixed 128-bit vector instructions). These systems are summarised in Table~\ref{tab:compute-system}. It should be noted that we recognise the A64FX processor is designed for HPC applications and completely different in nature to the RISC-V cores, which are designed for embedded and single-board computers (SBC). However, a comparison against the A64FX is still valuable as it can highlight important differences and potential design improvements for an HPC-class RISC-V processor in the future.

\begin{table}
\caption{Compute system specifications}\label{tab:compute-system}
\begin{adjustbox}{width={\textwidth},totalheight={\textheight},keepaspectratio}
\begin{tabular}{|>{\centering\arraybackslash}m{0.2\linewidth}|>{\centering\arraybackslash}m{0.3\linewidth}|>{\centering\arraybackslash}m{0.3\linewidth}||>{\centering\arraybackslash}m{0.3\linewidth}|}
\hline
 & \textbf{Allwinner D1} & \textbf{StarFive JH7110 (VF2)} & \textbf{A64FX}\\
 \hline \hline
 \textbf{Processor} & XuanTie C906 & SiFive U74 &Fujitsu A64FX\\
\textbf{ Clock speed }& 1.0GHz  & 1.5GHz & 1.8GHz\\
 \textbf{Cores} & 1  & 4 & 48\\
 \textbf{Cache} & 32 KB I-cache + 32 KB D-cache  & 32 KB I-cache + 32 KB D-cache + 2MB L2 & 64 KB I-cache + 64KB D-cache, 8 MB shared L2 cache per 12 cores (core memory group)\\
 \textbf{Memory} & 512MB DDR3  & 8GB DDR4 & 32GB HBM2\\
 \hline 
 \textbf{ISA} & RV64GC+V0.7  & RV64GC & ARMv8.2 with SVE\\
 \textbf{Vector width} & 128bit  & N/A & dual 128-bit (NEON) / dual 512-bit (SVE)\\
 \hline
\end{tabular}
\end{adjustbox}
\end{table}

\subsection{Methodology}
To evaluate the vectorisation performance we use the RAJA Performance Suite (RAJAPerf)~\cite{noauthor_llnlrajaperf_2023}, which comprises the following sets of benchmarks: ALGORITHM, APPS, BASIC, LCALS (Livermore Compiler Analysis Loop Suite), POLYBENCH, and STREAM (Babel Stream).
Since the C906 only supports vector element sizes up to 32-bit, we configure the benchmark to use the single-precision floating point data type. The compilers and respective compiler flags for RISC-V and Arm systems are specified in Table~\ref{tab:compiler-specs}. The benchmark timings are averaged over three runs.

\begin{table}
\caption{Compiler specifications}\label{tab:compiler-specs}
\begin{adjustbox}{width={\textwidth},totalheight={\textheight},keepaspectratio}
\begin{tabular}{|c|c|c|p{0.5\linewidth}|}
\hline
 \textbf{Name} & \textbf{Compiler } & \textbf{Vector width} & \textbf{Compiler flags} \\
\hline \hline
 RV-GCC8.4-scalar & XuanTie GCC 8.4 & N/A &{\tt -O3 -march=rv64gc -ffast-math} \\
 RV-GCC8.4-vector & XuanTie GCC 8.4 & 128-bit & {\tt  -O3 -march=rv64gcv0p7 -ffast-math} \\
 ARM-GCC11.2-scalar & GCC 11.2 & N/A & {\tt -O3 -ffast-math -mcpu=a64fx -march=armv8.2-a+nosimd+nosve}\\
 ARM-GCC11.2-vector & GCC 11.2 & 128-bit & {\tt -O3 -ffast-math -mcpu=a64fx -march=armv8.2-a+simd+nosve} \\ 

\hline
\end{tabular}
\end{adjustbox}
\end{table}

\subsection{Results}\label{results}
Table~\ref{tab:vectorized-kernels} summarises the list of kernels which are vectorised by the XuanTie GCC 8.4 compiler. It can be seen that 30 of the 64 kernels are successfully vectorised by the compiler, but for 7 of these only the scalar code and no vector instructions were executed at runtime. This is due to the compiler's oversensitivity to loop ranges, and the scalar branch is preferred and executed even when a vectorised branch is available. Clang 15.0, which generates RVV v1.0 assembly, is capable of vectorising more kernels than GCC 8.4; for a full comparison, see~\cite{jkll_riscv_vector_tool_2023}. 

\begin{table}[]
\caption{RAJA Performance Suite Kernels vectorised by RV-GCC8.4-vector}\label{tab:vectorized-kernels}
%\begin{adjustbox}{width={\textwidth},totalheight={\textheight},keepaspectratio}
\begin{tabular}{|>{\raggedright\arraybackslash}p{0.2\linewidth}>{\raggedright\arraybackslash}p{0.65\linewidth}r|}
\hline
\textbf{Kernels} & & \\ \hline \hline
\textbf{Vectorised and} & \textbf{executed} & \textbf{Total: 23}  \\ \hline
\textbf{Algorithm:} & MEMCPY, MEMSET, REDUCE\_SUM & \\ 
\textbf{Apps:} & ENERGY, FIR, PRESSURE & \\ 
\textbf{Basic:} & AXPY, AXPY\_ATOMIC, REDUCE3\_INT & \\ 
\textbf{Lcals:} & GEN\_LIN\_RECUR  & \\ 
\textbf{Polybench:} & 2MM, 3MM, ATAX, FDTD\_2D, GEMM, GEMVER, GESUMMV, MVT & \\ 
\textbf{Stream:} & ADD, COPY, DOT, MUL, TRIAD &  \\ \hline
\hline
\textbf{Vectorised} & & \textbf{Total: 7} \\ \hline
\textbf{Lcals:} & FIRST\_SUM, FIRST\_DIFF, HYDRO\_1D, HYDRO\_2D, TRIDIAG\_ELIM  & \\ 
\textbf{Polybench:} & JACOBI\_1D, JACOBI\_2D & \\ \hline 
\hline
\textbf{Scalar} & & \textbf{Total: 34}  \\ \hline
\textbf{Algorithm:} & SCAN, SORT, SORTPAIRS  &  \\ 
\textbf{Apps:} & CONVECTION3DPA, DEL\_DOT\_VEC\_2D, DIFFUSION3DPA, HALOEXCHANGE, HALOEXCHANGE\_FUSED, LTIMES, LTIMES\_NOVIEW, MASS3DPA, NODAL\_ACCUMULATION\_3D, VOL3D & \\ 
\textbf{Basic:} & IF\_QUAD, INDEXLIST, INDEXLIST\_3LOOP, INIT\_VIEW1D, INIT\_VIEW1D\_OFFSET, INIT3, MAT\_MAT\_SHARED, MULADDSUB, NESTED\_INIT, PI\_ATOMIC, PI\_REDUCE, REDUCE\_STRUCT, TRAP\_INT & \\ 
\textbf{Lcals:} & DIFF\_PREDICT, EOS, FIRST\_MIN, INT\_PREDICT, PLANCKIAN  & \\ 
\textbf{Polybench:} & ADI, FLOYD\_WARSHALL, HEAT\_3D &  \\ \hline
\end{tabular}
%\end{adjustbox}
\end{table}

\begin{figure}
\caption{Normalised runtime for RAJA Performance Suite kernels. ARM-GCC11.2-vector result (orange bars) are normalised against ARM-GCC11.2-scalar on A64FX, and both D1-RV-GCC8.4-vector (purple bars) and VF2-RV-GCC8.4-scalar (green bars) are normalised against D1-RV-GCC8.4-scalar.}\label{fig:normalized-runtime}
\centering
     \begin{subfigure}[b]{\textwidth}
         \centering
         \includegraphics[width=\textwidth]{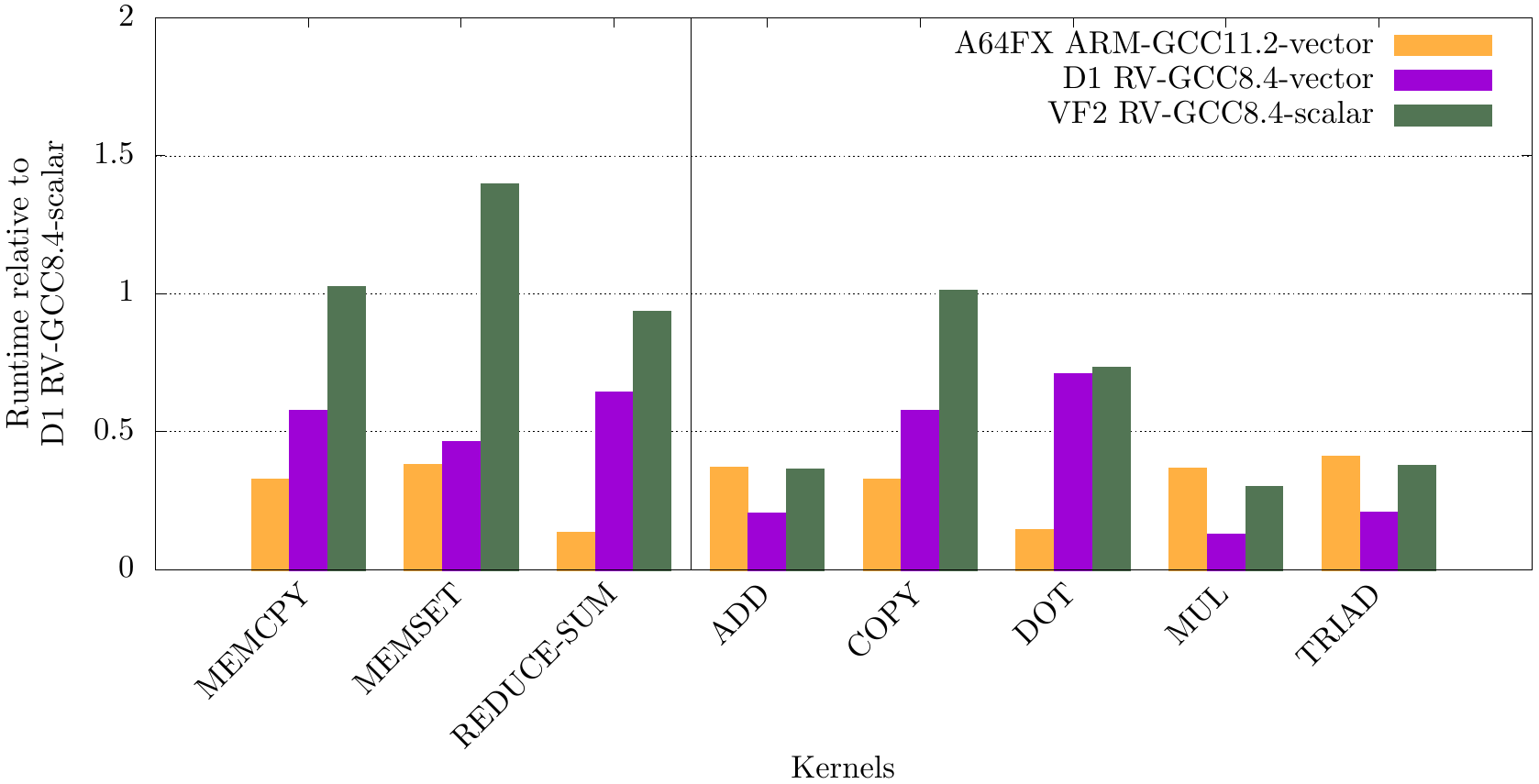}
         \caption{Algorithm (left) and Stream (right) Kernels}
         \label{fig:normalized-runtime-algorithm-stream}
     \end{subfigure}
      \begin{subfigure}[b]{\textwidth}
         \centering
         \includegraphics[width=\textwidth]{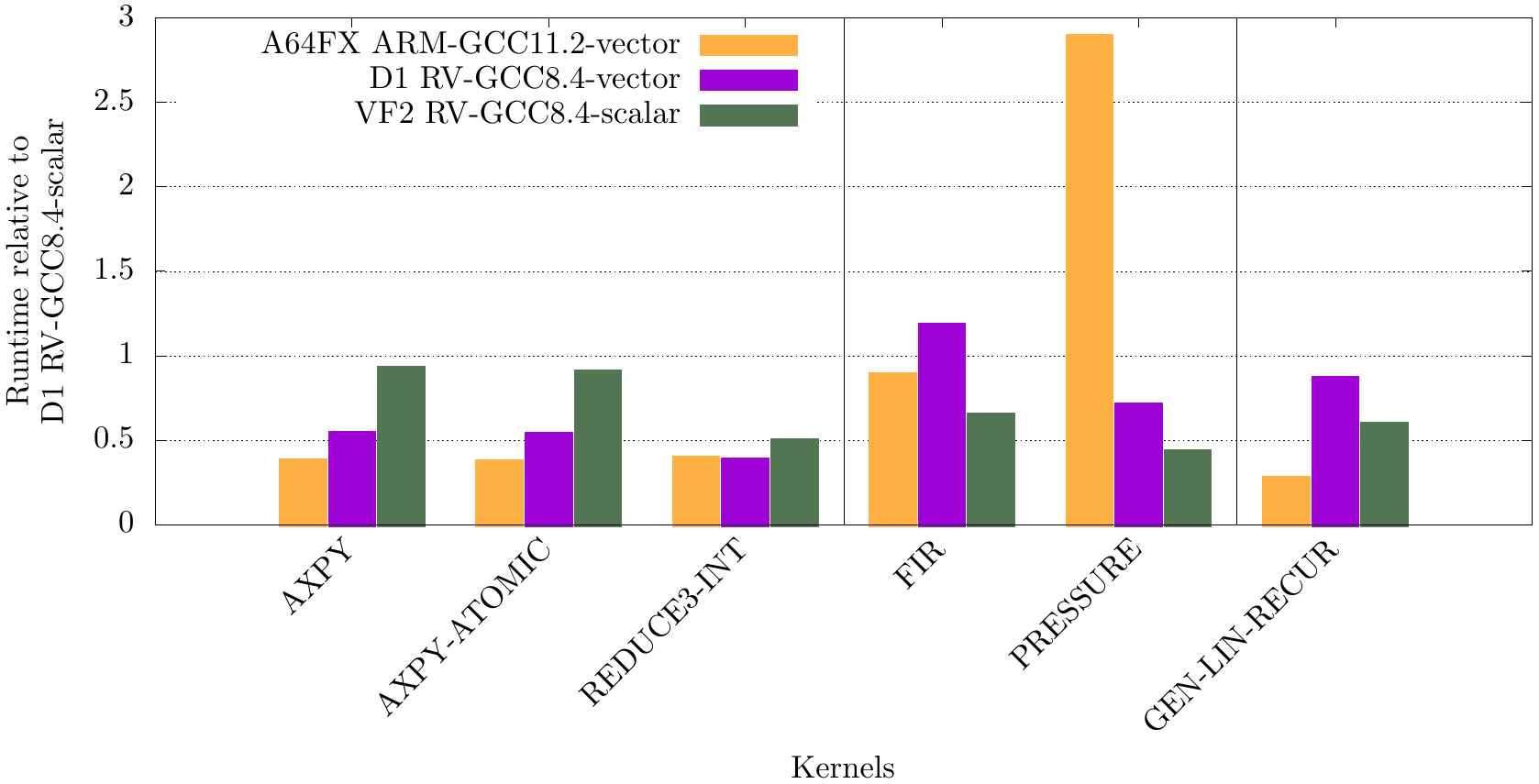}
         \caption{Basic (left), Apps (centre), and Lcals (right) Kernels}
         \label{fig:normalized-runtime-basic-apps-lcals}
     \end{subfigure}
     \begin{subfigure}[b]{\textwidth}
         \centering
         \includegraphics[width=\textwidth]{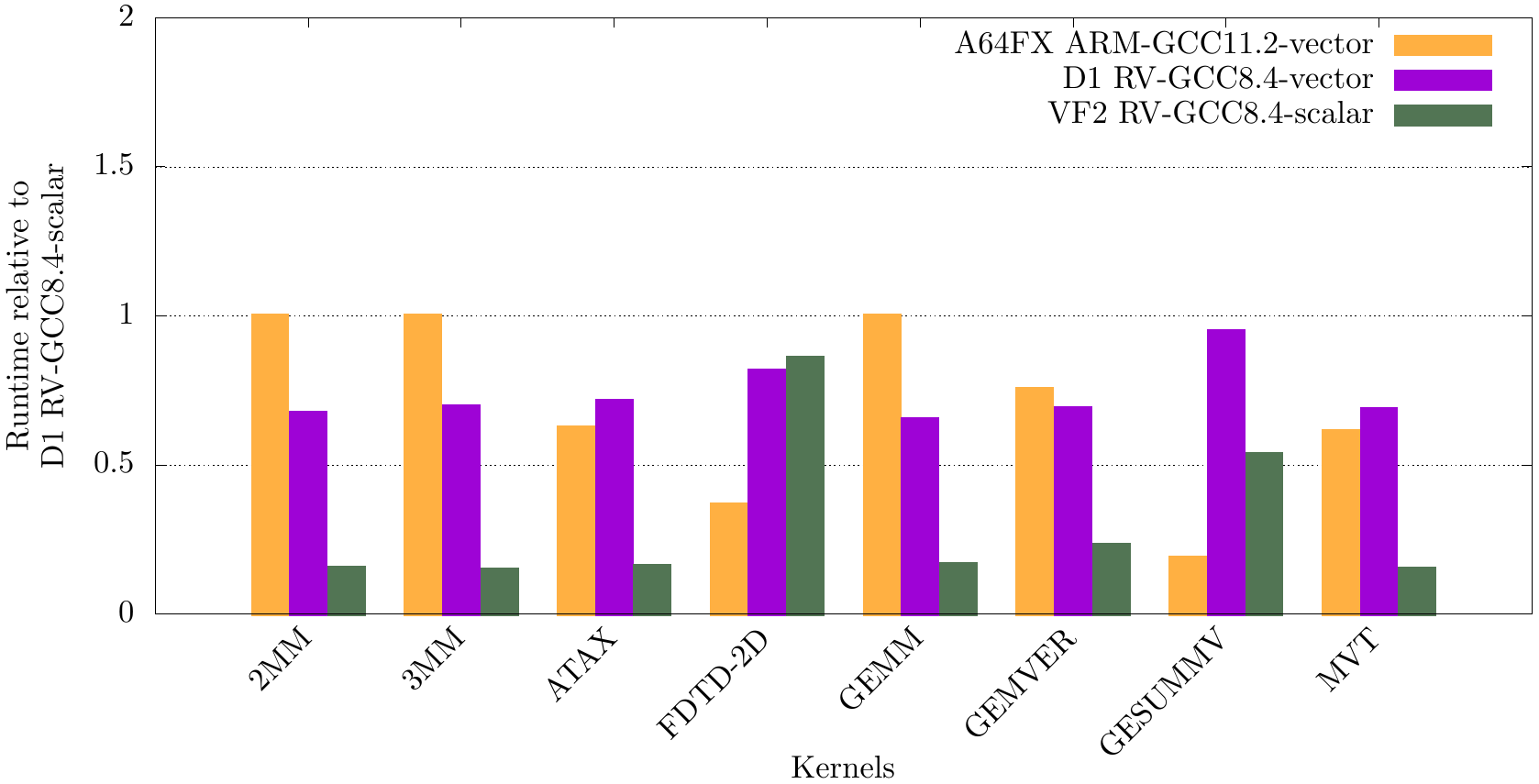}
         \caption{Polybench Kernels}
         \label{fig:normalized-rutime-polybench}
     \end{subfigure}
 \end{figure}

Figure~\ref{fig:normalized-runtime} reports runtimes for the RAJAPerf kernel normalised against the kernel's scalar runtime. For the A64FX, normalisation is against running in scalar mode on the A64FX, whereas for the Allwinner D1 and StarFive JH7110 it is normalised against running scalar on the D1. The orange and purple bars show the vectorisation performance difference on the A64FX and D1 respectively, and the green bars show a comparison of the scalar performance between the JH7110 (VF2) and the D1.

It can be observed from these plots that for most linear algebra kernels, the vectorised code on the RISC-V D1 is faster compared to its scalar counterpart, at around 84\% faster for AXPY, 53\% for GEMM, 45\% for GEMVER, 40\% for ATAX, and 46\% for MVT. Vectorised code also sustain much higher bandwidth for streaming kernels such as Stream ADD, COPY, DOT, MUL, and TRIAD. In only one case, the FIR kernel, is the vectorised code slower than its scalar counterpart. Whilst in most cases the speedup from RVV on the D1 is not as significant as from NEON on A64FX, there are some exceptions; for example, matrix multiplication kernels on the A64FX compiled with ARM-GCC11.2-vector did not execute the vector instructions. Therefore, the runtime performance was the same as the scalar executable. Furthermore, the vectorised A64FX PRESSURE kernel was almost three times slower than the scalar version.

When comparing the RISC-V processors AllWinner D1 and StarFive JH7110, it can be observed that for high arithmetic intensity kernels the JH7110 (VF2), which has a higher clock frequency, significantly outperforms the D1. For example, GEMM is six times faster on the VF2 compared to running scalar on D1, and four times faster than the vectorised version of this benchmark on the D1. However, even though the theoretic memory bandwidth for the VF2 is higher than the D1, these benchmarking results demonstrate that with vectorisation the D1 executes the streaming kernels faster than the VF2. For example, Stream ADD is 82\% faster and COPY is 77\% faster on the D1. This is the reason why we observe that the D1 can perform low arithmetic intensity operations faster than VF2, for example AXPY on D1 with vectorisation enabled is 71\% faster than the VF2 which is running in scalar mode.

\ifdefined\Outlook
\section{Outlook}

\fbox{\parbox{0.9\textwidth}{\textbf{MJ:} I think we should include the roadmap for new RISC-V boards e.g. 4-core C910 less RVV }}
\fi

\section{Conclusions and recommendations}
At the time of writing, generating and testing RVV codes on the currently available physical CPUs is problematic due to the mismatch between the available tooling, such as GCC and Clang, and the RVV version (v0.7.1) implemented in hardware. However, as demonstrated in Section \ref{results}, compiling for RVV on the D1 can result in codes being up to 80\% faster than the scalar alternative (RAJAPerf AXPY and Stream ADD). The standardisation of tooling with v1.0 RVV and intrinsics will greatly simplify the development of vectorised codes in the future, running on RVV v1.0 compliant CPUs. Therefore our view is that, whilst at the time of writing there are challenges around developing and running vectorised code on RISC-V due to the immaturity of tooling and hardware, in the medium term these challenges will be solved and RVV provides a strong foundation for leveraging RISC-V for high performance workloads. Furthermore, the improved auto-vectorisation of LLVM, coupled with increased VLEN in future CPUs, is expected to increase kernel runtime performance even further.

Although the later versions of the T-Head GCC toolchain supports both RVV v0.7 and v1.0, neither the mainstream GCC or LLVM toolchains support v0.7. Whilst it is understandable that the toolchain development teams only want to support the ratified version of RVV, the currently available RVV hard CPU cores only support v0.7 and the runtime performance benefits of leveraging RVV on the C906-based devices are tangible, as shown in Section \ref{results}. Furthermore, T-Head have proven that it is possible to provide RVV v0.7 and RVV v1.0 support within the GCC toolchain, providing the \texttt{-march=rv64gcv0p7} and \texttt{-march=rv64gcv1p0} compiler options. With the large volume of RVV v0.7 devices in circulation we would like to see support for both v0.7 and v1.0 RVV in mainstream GCC and Clang / LLVM toolchains.

\subsection{Recommendations}
In order to leverage the runtime performance benefits of vectorisation on current RISC-V hardware and to minimise the impact of the code incompatibilities between RVV v0.7 and v1.0 \cite{hsiangkai_wang_risc-v_2021}, we recommend the use of the T-Head GCC 8.4 auto-vectorisation and not using the T-Head RVV v0.7 intrinsic API. This will ensure that codes can simply be recompiled, without modification, to target RVV v1.0 compatible hardware. Another option, is to generate code for RVV v1.0 using GCC or Clang / LLVM auto-vectorisation or the v1.0 intrinsics API, and utilise a conversion tool such as  \cite{jkll_riscv_vector_tool_2023} to create binaries for RVV v0.7 hardware.

We would also recommend building RVV-enabled Linux images with a patched mainstream \emph{buildroot} using the T-Head GCC 8.4 compiler, as support for the Allwinner D1 has recently been added.

\section{Acknowledgement}
The authors would like to thank the ExCALIBUR H\&ES RISC-V testbed for access to compute resource used in this work.

 \bibliographystyle{splncs04}
 \bibliography{reference}
%\end{thebibliography}
\end{document}